\documentclass[pre,twocolumn,showpacs]{revtex4}
\usepackage{amssymb,amsmath}
\usepackage{epsfig}
\usepackage{graphicx}
\usepackage{latexsym}

\newcommand{\bit}{\begin{itemize}}
\newcommand{\eit}{\end{itemize}}
\newcommand{\beq}{\begin{equation}}
\newcommand{\eeq}{\end{equation}}
\newcommand{\bea}{\begin{eqnarray}}
\newcommand{\eea}{\end{eqnarray}}

\begin{document}
\title{Rate dependent shear bands in a shear transformation zone model of amorphous solids}
\author{M. L. Manning}%
 \email{mmanning@physics.ucsb.edu}
\affiliation{%
Center for Theoretical Science, Princeton University, NJ 08544
}%
\affiliation{%
Department of Physics, University of California, Santa Barbara, 93106
}%
\author{E. G. Daub}
\author{J. S. Langer}
\author{J. M. Carlson}
\affiliation{%
Department of Physics, University of California, Santa Barbara, 93106
}%
\date{\today}

\begin{abstract}
  We use Shear Transformation Zone (STZ) theory to develop a deformation map for amorphous solids as a function of the imposed shear rate and initial material preparation. The STZ formulation incorporates recent simulation results [Haxton and Liu, PRL {\bf 99} 195701 (2007)] showing that the steady state effective temperature is rate dependent. The resulting model predicts a wide range of deformation behavior as a function of the initial conditions, including homogeneous deformation, broad shear bands, extremely thin shear bands, and the onset of material failure. In particular, the STZ model predicts homogeneous deformation for shorter quench times and lower strain rates, and inhomogeneous deformation for longer quench times and higher strain rates. The location of the transition between homogeneous and inhomogeneous flow on the deformation map is determined in part by the steady state effective temperature, which is likely material dependent. This model also suggests that material failure occurs due to a runaway feedback between shear heating and the local disorder, and provides an explanation for the thickness of shear bands near the onset of material failure. We find that this model, which resolves dynamics within a sheared material interface, predicts that the stress weakens with strain much more rapidly than a similar model which uses a single state variable to specify internal dynamics on the interface. 
\end{abstract}

\maketitle

\section{Introduction}
\label{RDbg}
 
  Amorphous solids such as foams, dense colloids, bulk metallic glasses, and granular fault gouge are ubiquitous in engineering applications and natural systems. Although these materials exhibit a yield stress on experimental time scales, they flow, deform, and fail in a manner which is different from crystalline solids or Newtonian fluids. Many of these materials undergo strain localization, where a small region deforms much more rapidly than adjacent regions. For example, bulk metallic glasses develop very thin shear bands~\cite{Johnson2, Johnson3}, fault gouge in earthquake faults develops a prominent fracture surface that accommodates most of the slip~\cite{Chester}, and colloidal systems develop broad shear bands~\cite{Cohen}.

  Surprisingly, the mechanisms that lead to this strain localization have remained elusive. An early theory of shear banding~\cite{Griggs} suggests that a small increase in the thermal temperature lowers the viscosity, resulting in more rapid deformation and a local increase in temperature.  However, Lewandowski and Greer showed that shear bands in bulk metallic glasses can not be explained by adiabatic thermal effects~\cite{Lewandowski}. Although thermal heating must play a role at high strain rates, it does not appear to govern the formation of shear bands in many materials.

  In a recent paper~\cite{MANNING-LANGER}, we found that at low strain rates the Shear Transformation Zone (STZ) theory for amorphous solids predicts shear band formation. These bands are generated by feedback between the local strain rate and the configurational disorder of an amorphous packing. In a separate paper ~\cite{LANGER-MANNING}, we used STZ theory to fit data over a wide range of strain rates from a simulation of glassy disks by Haxton and Liu~\cite{HL07}. These data indicate that dynamics of the glassy material change dramatically at large strain rates.

 Experimental studies of bulk metallic glasses driven at a wide range of strain rates also show that deformation is strain rate dependent.  Homogeneous deformation is seen at low strain rates and inhomogeneous flows dominate at large strain rates~\cite{Jiang, Lu}.

 In this paper, we study deformation in the STZ model as a function of the initial material preparation and externally imposed strain rate, and include the rate dependence observed by Haxton and Liu in our model. This STZ formulation is valid for a large range of strain rates and predicts four different types of deformation behavior: homogeneous deformation, thick disorder limited shear bands, thin diffusion limited shear bands, and material failure. We will discuss these various types of deformation in detail in the following sections. We describe the mechanism that generates shear bands and the processes that determine the thickness and longevity of these inhomogeneous flows. We numerically integrate the STZ equations to produce a deformation map for the model glass simulated by Haxton and Liu that shows the type of deformation or failure predicted as a function of the initial conditions.

  Shear bands in this STZ formulation occur due to a feedback between the effective temperature, which describes the configurational disorder in a glassy or jammed material, and the local strain rate. In a sheared, steady state, non equilibrium amorphous material, the effective temperature can be calculated by measuring the fluctuations and linear response of an observable such as the pressure and applying the fluctuation-dissipation theorem (FDT)~\cite{Cugliandolo, Ono, HL07}. Ono, {\em et al.} have shown that in several simulated foams, measurements of different observables yield a single, rate dependent steady state effective temperature which is distinct from the thermal temperature~\cite{Ono}. In addition, these authors show that the FDT effective temperature is consistent with an entropic definition: the effective temperature is the derivative of the configurational entropy with respect to the (potential) energy. This definition suggests that slow steady shearing causes the material to ergodically explore all possible configurational packings, and therefore the system maximizes a configurational entropy.

  The STZ model describes plastic deformation in amorphous material as a series of localized plastic events that occur in susceptible regions, or zones~\cite{Argon,Bulatov,Falk_L1}. Following Falk and Langer, we model STZs as bistable regions that are more likely than the surrounding material to deform under stress, and are created and annihilated as energy is dissipated in the system~\cite{Falk_L1}. This model has successfully been used to describe bulk metallic glasses, thin films, and hard spheres in several different geometries~\cite{MANNING-LANGER,JSL08,Gregg,Anael1, Falk_band}.

 An important feature of all STZ formulations is that the zones are activated by an effective temperature or free volume, and there is a feedback between packing structure and deformation. In particular, we postulate that an STZ is an unlikely, high energy configuration of an amorphous packing.  Because the effective temperature governs the statistics of configurational packings, the steady state density of STZs should correspond to a Boltzmann factor,
\begin{equation}
\hat{\Lambda} = \exp \left[\frac{- E_z}{ k_B \, T_{eff}} \right] \equiv \; \exp \left[-\frac{1 }{ \chi} \right],
\end{equation}
where $\hat{\Lambda}$ is proportional to the steady state STZ density, $E_z$ is an activation energy, $k_B$ is the Boltzmann constant, $T_{eff}$ is the effective temperature, and $\chi \equiv E_z/ (k_B T_{eff})$ is a dimensionless effective temperature~\cite{JSL08}.

  Because plastic deformation occurs only at these STZs, the plastic strain rate in simple shear is proportional to this density, and in many situations can be written as follows:
\begin{equation}
\label{RDq-chi}
\tau_0 \dot\gamma_{pl} = 2 \; e^{-1/\chi} f(s),
\end{equation}
where $\gamma$ is strain, $f(s)$ is a function of the deviatoric stress $s$ and $\tau_0$ is an internal timescale such as the phonon frequency.  For the remainder of this paper we will refer to the dimensionless plastic strain rate, $q = \dot\gamma_{pl} \tau_0$ and $\overline{q} = (V_0/ L) \, \tau_0$ is the imposed average strain rate times the STZ time scale.
 
For completeness, Appendix~\ref{sec:RDmodel_detail} reviews the STZ model and defines the exact equations and parameters used in this work. The model is vastly simplified by focusing on materials in a simple shear geometry at thermal temperatures far below the glass transition temperature, and can be summarized by two equations.

   The first specifies that elastic deformation increases the shear stress, while plastic deformation decreases it:
\begin{equation}
\label{dot-s}
\frac{ds}{d \gamma} = \mu_{*} \left( 1 - \frac{2 }{\overline{q}} \, f(s)\, \overline{\Lambda} \right),
\end{equation}
where $\mu_{*}$ is the ratio of the elastic modulus to the yield stress, and $\overline{\Lambda}$ is the spatial average of the STZ density $\Lambda = \exp(-1/ \chi)$. In writing this equation we have assumed that the deviatoric stress is constant across the width of the sample, as discussed in the Appendix.

  A second equation specifies that the effective temperature approaches its steady state value, $\hat\chi(q)$ as plastic work is dissipated, and it also diffuses:
\begin{eqnarray}
\label{dot-chi}
\frac{ d \chi}{d \gamma} &=& \frac{2  \: s \chi}{\tilde{c}_0 s_0 \overline{q}}f(s)  e^{-1/ \chi}  ( 1 - \frac{\chi}{\hat{\chi}(q)}) + \:  a^2 \dot{\gamma}_{pl}  \frac{\partial^2 \chi}{\partial y^2},
\end{eqnarray}
where  $c_0$ is a dimensionless specific heat, $a$ is a diffusion length scale on the order of an STZ radius, and $s_0$ is the stress threshold for the onset of plastic deformation. For the remainder of this paper, a symbol with an overline denotes a quantity that is a spatial average or constant as a function of position.  The function $f(s)$ is described in the Appendix.  

 The remaining unspecified component of Eq.~\ref{dot-chi} is the steady state effective temperature $\hat{\chi}$. This material dependent parameter captures the physically intuitive idea that there is a maximum possible disorder attained in a sheared amorphous packing; above $\hat\chi$ no heat can be dissipated in the configurational degrees of freedom.  Recent simulations by Haxton and Liu~\cite{HL07} suggest that the steady state effective temperature is dependent on the plastic strain rate, denoted $q$.  In steady state, the inverse strain rate $1/q$ can be viewed as a function of the steady state effective temperature $\hat{\chi}$, much like the viscosity is a function of the thermal temperature $T$. We have shown~\cite{LANGER-MANNING} that the the glassy steady state effective temperature is well fit by the following functional form:
\begin{equation}
\label{chi-hatI}
\frac{1}{q(\hat{\chi})} = \frac{1}{q_0}\exp \left[ \frac{A}{\hat{\chi}} + \alpha_{eff}(\hat{\chi}) \right],
\end{equation}
A discussion of this effective temperature glass transition is given in~\cite{LANGER-MANNING}, we adopt the super-Arrhenius function $\alpha_{eff}$ identified in that paper:
\begin{equation}
\label{alpha_effRD}
\alpha_{eff}(\chi) = \frac{\chi_1}{\chi- \chi_0}\exp[ - b (\chi- \chi_0)/(\chi_A-\chi_0)],
\end{equation}
where $\chi_0$ is the thermal glass transition temperature, $\chi_A$ is the temperature at which the system begins to exhibit Arrhenius behavior, and $b$, $\chi_1$ are fit parameters. The super-Arrhenius component (Eq.~(\ref{alpha_effRD})) ensures that the effective temperature approaches a constant, $\chi_0$, as the strain rate approaches zero, which is seen in simulations.

   As noted above, this paper focuses on deformation in geometries where the equilibrium shear stress is constant across the sample. Although there are many interesting geometries with shear stress gradients (and STZ theory can explain these shear bands~\cite{Falk_band}), often in these cases stress effects compete with internal structural effects and complicate the analysis. In contrast, experiments in simple shear geometries exhibit strain localization even though the equilibrium stress is constant in space, indicating that some property of the internal state governs shear band formation. 

  Shear banding is caused by the coupling in Eq.~(\ref{RDq-chi}) between a configurational or structural parameter and the plastic rate of deformation.  Although the stress equilibrates quickly, the effective temperature dynamics continue to evolve over much longer timescales.  Materials that develop long-lived shear bands exhibit a very different macroscopic rheology compared to those that deform homogeneously~\cite{Shi}. The type of deformation has important implications for the macroscopic material and frictional properties --- systems that localize also weaken rapidly.

   The remainder of this paper is organized as follows.  In Section~\ref{sec:RDstab} we study the stability of the model given by Eqs.~(\ref{dot-s}) and~(\ref{dot-chi}) with respect to perturbations. We numerically integrate the STZ equations to validate our analytic stability results in Section~\ref{sec:RDnum_res} and study the different types of deformation that persist for long times in Section~\ref{sec:RDlocal_dyn}. Section~\ref{conclusionsRD} concludes with a discussion of our results and open questions.

\section{Stability Analysis}
\label{sec:RDstab} 
   We now study shear band formation for systems with a rate dependent steady state effective temperature $\hat\chi(q)$. The fact that shear bands persist in the STZ model is somewhat surprising, given that the only stationary state of Eqs.~(\ref{dot-s}) and~(\ref{dot-chi}) is homogeneous deformation. In the following sections, we provide an explanation for shear band formation and evolution across a wide range of strain rates.

 We emphasize that all of the shear bands discussed below are transient phenomena -- at very large strains the shear bands diffuse across the entire width of the material and deformation once again becomes homogeneous.  However, because the internal dynamics of the configurational degrees of freedom can be very slow compared to the stress evolution timescale, shear bands persist for very long times. For example, the shear stress in numerical solutions to the STZ equations generally appears to reach a steady state within a few percent  strain but the effective temperature profile often remains inhomogenous for more than 20\% strain. 

First, the evolution equations for $\chi$ and $s$ can be written as follows:
\begin{eqnarray}
\label{s-dotRD}
\dot{s}(s, \chi) &=& \mu_{*} \left( 1 - 2 \, \frac{f(s)}{\overline{q}} \int  dy\, e^{-1/\chi} \right) ; \\
\label{chi-dotRD}
\dot{\chi}(s, \chi) &=&  2 f(s) e^{-1/\chi} \left( \frac{s \chi}{\overline{q} c_0 s_0}\left(1 - \frac{\chi}{\hat{\chi}(q)}\right) + a^2 \frac{\partial^2 \chi}{\partial y^2} \right), 
\end{eqnarray}
where the $(\cdot)$ operator indicates a derivative with respect to time in units of strain. The function $\hat{\chi}(q) = \hat{\chi}( 2 f(s) e^{-1/\chi})$ depends implicitly on the stress and the effective temperature. Then the Jacobian $J$ is given by:
\begin{eqnarray}
J_{11} &=& \frac{d \dot{s}}{ds} =  \frac{ -2 \mu_{*}}{\overline{q}} e^{-1/\chi} f^{\prime}(s), \\
J_{12} &=& \frac{d \dot{s}}{d \chi} = \frac{ -2 \mu_{*}}{\overline{q}} e^{-1/\chi} f(s)W(\delta \chi) / \chi^2, \\
J_{21} &=&\frac{d \dot{\chi}}{ds} =  \frac{2\chi  e^{-1/\chi}}{s_0 c_0 \overline{q}} \nonumber \times \\
 & & \left[\left( 1 - \frac{\chi}{\hat{\chi}}\right) \left( s f^{\prime}(s) + f(s) \right) + \frac{\chi s f}{\hat{\chi}^2} \frac{\partial \hat{\chi}}{\partial s} \right],\\
J_{22} &=&\frac{d \dot{\chi}}{d \chi} =  \frac{2 e^{-1/\chi}f(s)s}{s_0 c_0 \overline{q} } \times \nonumber \\
& &  \left[ \left( 1 -\frac{\chi}{\hat{\chi}} \right) \left( 1 + \frac{1}{\chi} \right) +  \left( -\frac{\chi}{\hat{\chi}} + \frac{\chi^2}{\hat{\chi}^{2}} \frac{\partial \hat{\chi}}{\partial \chi} \right) \right] .
\end{eqnarray}
The term $W(\delta \chi)$ is a spatial integral over one period of the perturbation function; it selects only the zero wave number component of the perturbing function because the other components must satisfy periodic boundary conditions:
\begin{eqnarray}
W(\delta \chi) &=& \sum_{k = - \infty}^{\infty} \frac{1}{2L} \int_{-L}^{+L} d y \; \delta \chi_{k} \, e^{i k \pi y / L} ;\\
  &= &  \begin{cases} 1 & \mbox{for}  \; k = 0, \\ 
           0  & \mbox{for} \; k \neq 0. \end{cases}
\end{eqnarray}
 Details can be found in~\cite{MANNING-LANGER}. We will use the term ``zero-mean perturbation'' to refer to a perturbing function with a vanishing $k=0$ component that does not change the average value of the underlying function across the width of the sample.

  For a specified externally imposed strain rate $\overline{q} = \tau_0 V_0/ L $, the steady state solution to Eqs. \ref{s-dotRD} and \ref{chi-dotRD} is $\chi = \hat{\chi}(\overline{q})$, and $s_{ss}$ is given by the solution to the algebraic equation $1 = 2 f(s_{ss}) \exp[-1/ \hat{\chi}(\overline{q}) ] / \overline{q}$.  The Jacobian evaluated at this steady state is greatly simplified:
\begin{eqnarray}
J_{11} &=&  \frac{ -2 \mu_{*}}{\overline{q}} e^{-1/\chi}\, f^{\prime}(s), \\
J_{12} &=& \frac{- \mu_{*} \, W(\delta \chi) }{ \chi^2}, \\
J_{21} &=& \frac{s}{c_0 s_0} \frac{\partial \hat{\chi}}{\partial{s}},\\
J_{22} &=& \frac{s}{c_0 s_0} \left( \frac{\partial \hat{\chi}}{\partial{\chi}} - 1 \right).
\end{eqnarray}

The upper right entry ($J_{12}$) vanishes for zero-mean perturbations, so Jacobian is lower triangular, and the diagonal values alone specify stability. $J_{11}$ is strictly negative because $f$ is a monotonically increasing function of $s$, and therefore the system is linearly stable with respect to zero-mean perturbations from a homogeneous steady state if $J_{22}$ is negative, or equivalently, $\partial \hat{\chi}/ \partial{\chi} < 1$.

The partial derivative $\partial \hat{\chi}/ \partial{\chi}$ can be written as follows:
\begin{equation}
\label{dchi}
\frac{\partial \hat{\chi}}{\partial \chi} = \frac{\hat{\chi}'(q) q}{\chi^2}, 
\end{equation} 
where $\hat{\chi}'(q)$ is determined by the derivative of Eq.~\ref{chi-hatI}.

 This clarifies the physical meaning of the criterion $\partial \hat{\chi}/ \partial{\chi} < 1$. At low strain rates the super-Arrhenius function $\alpha_{eff}$ ensures that $\hat{\chi}$ changes slowly with $q$ and the partial derivative in Eq.~\ref{dchi} is significantly smaller than unity.  In~\cite{MANNING-LANGER} we used the approximation that $\hat{\chi}(q)$ is constant at low strain rates; in this case $\partial \hat{\chi}/ \partial{\chi}$ is trivially zero and the STZ equations are always stable with respect to perturbations in steady state.

 At high strain rates, where $\alpha_{ef\!f}$ is zero and $\hat{\chi}(q) = A/ \ln(q_0/q)$, it can be shown that $\partial \hat{\chi}/ \partial{\chi} = 1/ A$. In this regime, the equations are stable with respect to perturbations in steady state if and only if the normalized activation energy $A$ is greater than unity.

  The material dependent parameter $A$ is greater than unity in all homogeneously deforming materials where the steady state stress increases with increasing imposed strain rate.  To see this, note that  Eq.~\ref{chi-hatI} can be rewritten as follows:
\begin{equation}
\label{chi-q}
\hat{\chi}(\overline{q}) = - \frac{A_{*}(\overline{q})}{\ln(\overline{q}/ q_{*})},
\end{equation}
where $A_*$ is an activation energy and $q_{*}$ is a dimensionless constant. In steady state, $\chi=\hat{\chi}$ everywhere, so inserting Eq.~\ref{chi-q} into Eq.~\ref{RDq-chi} and taking the derivative with respect to $q$ results in the following equation for $s'(\overline{q}) $, which is the derivative of the steady state stress with respect to the imposed driving rate:
\begin{eqnarray}
\label{sprime}
s^\prime(q) &=&  \left[ -1 + A_*(q) + q \ln(q/q_{*})A_*^\prime (q))\right] \times \nonumber \\ 
& & \left( \frac{q}{q_{*}} \right)^{-1/A_{*}(q)} \frac{1}{ A_*(q) f^{\prime}(s)}
\end{eqnarray}

The function $f$ increases monotonically with $s$. Therefore $s'(q)$ is positive and the steady state stress increases with increasing imposed strain rate whenever the first factor in Eq.~\ref{sprime} is positive.  In high strain regimes, the activation energy is a constant,  $A_*(q) = A$, and the material is steady state strengthening whenever $A > 1$.

 In most simulations of disordered disks with hard core repulsion (including simulation data by Haxton and Liu~\cite{HL07} and Shi {\em et al.}~\cite{Shi}), the steady state stress increases with increasing strain rate, so that $\partial \hat{\chi} / \partial{\chi}$ is less than unity and the system is stable with respect to zero mean perturbations. Materials with a steady state stress that increases as a function of driving rate are called rate strengthening.  In contrast, several experiments that study friction of granular fault gouge find rate weakening behavior; the steady state stress decreases as the imposed strain rate increases~\cite{Marone_review}.  These materials might be susceptible to shear banding even in steady state.  In this paper we focus on shear banding in materials that are rate strengthening, but rate weakening materials are an interesting avenue of further research.

   STZ theory predicts that rate strengthening materials should continue to deform homogeneously once they attain steady state. However, for certain initial conditions simulated rate strengthening materials develop shear bands before they reach this homogeneous steady state. This leads us to study the stability of perturbations at each point in time along a time-varying trajectory, assuming that the system deforms homogeneously up to that point. This is a ``frozen time'' stability analysis.

 Due to the integral in Eq.~(\ref{s-dotRD}), spatial perturbations to $\chi$ with zero mean do not change the equation for $\dot{s}$ to first order in $\delta \chi$ : $\dot{s}(s, \overline{\chi} + \delta \chi)= \dot{s}(s, \overline{\chi}) + \mathcal{O}( \delta \chi^2)$. In this case the linear stability is determined solely by the equation for $\chi$ and the sign of $J_{22}$. Above the yield stress, $\chi$ and $f(s)$ are strictly greater than zero and therefore $J_{22}$ is negative (and the trajectory is linearly stable) whenever the following criterion is met:

\begin{eqnarray}
\label{chi_crit}
\chi > \chi_{crit} &=& \frac{1}{4 \hat \chi}\left( -\hat{\chi} + \hat{\chi}^2 + q \hat{\chi}'(q) + \right. \nonumber \\
& &\left. \sqrt{8 \hat{\chi}^3 + (- \hat{\chi} + \hat{\chi}^2 + q \hat{\chi}'(q))^2} \right).
\end{eqnarray}

 By inserting values for $\hat\chi(q)$ into Eq.~(\ref{chi_crit}), we derive a linear stability prediction for the boundary between these two regimes as a function of the average initial effective temperature $\overline\chi$ and the applied strain rate $q$. We choose $\hat\chi(q)$ to fit the data from Haxton and Liu~\cite{HL07,LANGER-MANNING}.

\begin{figure}[h!]
\centering \includegraphics[height=7cm]{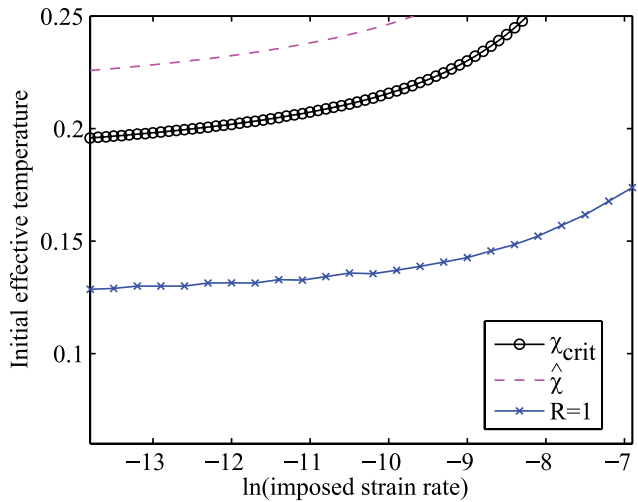}
\caption{\label{fig:band_phase} (color online) Predicted deformation map based on initial conditions only. The solid line marked with circles (black) represents the frozen time linear stability criterion predicted by Eq.~(\ref{chi_crit}), which does not take into account finite amplitude perturbations.  $\mathcal{R} $ is a more accurate generalized stability criterion that takes into account finite amplitude perturbations. A line with constant localization ratio $\mathcal{R}=1$ is marked with crosses(blue).  Above the line,  $\mathcal{R} < 1$ and homogeneous deformation is predicted, while below the line $\mathcal{R} > 1$ and strain localization is predicted. For reference, the upper dashed line shows the $\hat\chi(q)$ fit to the data from Haxton and Liu~\cite{HL07}.}
\end{figure}

 Unlike linear stability analysis for steady states, frozen time stability analysis for time varying trajectories does not predict the final state of the system. It provides an indication that a transient instability is possible, but it does not specify global stability. The frozen time analysis is accurate only when the perturbations grow rapidly compared to the growth of the underlying trajectory.  We therefore use the more general localization ratio $\mathcal{R}$ to characterize the transient instability. First discussed in ~\cite{MANNING-LANGER},  this ratio compares the growth rate of perturbations (determined by frozen-time stability analysis) to the growth of the underlying trajectory:
\begin{equation}
\label{locR}
\mathcal{R} = \delta \chi \frac{\exp[J_{22}(s_m, \chi_{ini}) \Delta t/2 ] J_{22}(s_m, \chi)}{\dot{\chi}(s_m,\chi)},
\end{equation}
where $\delta \chi$ is the magnitude of the perturbation, $\chi_{ini}$ is the initial effective temperature, $s_m$ is the approximate maximum shear stress given by the solution of the equation $\overline{q} = 2  f(s_m) \;\exp[-1/\chi_{ini}] $, and $\Delta t$ is the approximate time in units of strain it takes to achieve the stress maximum. Localization occurs when the rate at which heat is dissipated inside the band is larger than the rate outside the band; in this case $\mathcal{R}$ is greater than unity.

  The localization ratio given by Eq.~(\ref{locR}) depends on the magnitude of the perturbation $\delta \chi$. In the low strain rate limit, we have systematically studied the localization ratio $\mathcal{R}$ as a function of perturbation amplitudes and found that it accurately predicts that larger perturbations lead to enhanced localization~\cite{MANNING-LANGER}.  For simplicity, we have chosen $\delta \chi$ to be 5 \% of the average value of the effective temperature, which is consistent with perturbations to the potential energy per atom for a Lennard Jones glass calculated by Shi, {\em et al.}~\cite{Shi}. Systematically studying the effects of perturbation magnitude as a function of strain rate is beyond the scope of this paper.

     Figure~\ref{fig:band_phase} is a deformation map that predicts the type of flow as a function of the initial conditions for the simulated glassy material studied by Haxton and Liu~\cite{HL07}.  The bold line is the linear stability criterion defined by Eq.~(\ref{chi_crit}). Because a frozen time analysis does not take into account finite amplitude perturbations or the growth rate of the underlying trajectory, we use the localization ratio, $\mathcal{R}$, to predict localization. Using Eq.~\ref{locR}, we calculate $\mathcal{R}$ for each set of initial conditions with $\delta \chi = 0.05 \times \chi_{ini}$, and $\Delta t = 0.06$= 6~\% strain.  The line marked with crosses in Fig.~\ref{fig:band_phase} corresponds to a line with constant $\mathcal{R} = 1$. Localization is expected below this line, where $\mathcal{R} > 1$, and homogeneous flow above it.   

\section{Numerical solutions to STZ equations}
\label{sec:RDnum_res}

  To check these analytic predictions, we numerically integrate the STZ partial differential equations. The numerical solutions exhibit three broad categories of deformation behavior: homogeneous deformation, shear bands, and melting or failure.  This section discusses qualitative features of each kind of deformation, while Section~\ref{sec:RDlocal_dyn} develops a deformation map using a quantitative criterion for each category and discusses macroscopic implications.

 To resolve extreme localization, we use an irregular mesh and a combination of fixed-step and adaptive-step finite difference methods. For each pair of initial conditions, the average initial effective temperature $\chi_{ini}$ and the externally applied strain rate $\overline{q} = \tau_0 (V_0/ L)$, we numerically integrate the STZ equations (Eqs.~(\ref{dot-s}) and~(\ref{dot-chi})) from 0 to 20 \% strain.  The initial effective temperature function $\chi_{ini}(y)$ is a constant perturbed by a hyperbolic secant function of height $\delta \chi$ and width $L/10$, normalized so that its average is $\chi_{ini}$, and the initial shear stress is $0.0001$. All stresses are in units of the yield stress $s_y$ unless otherwise noted.

  For comparison, we also numerically integrate a single degree of freedom STZ model, where the effective temperature is constrained to be constant across the width of the material, and no perturbations are permitted. The system of ordinary differential equations given by Eqs.~(\ref{dot-s}) and~(\ref{dot-chi}) (with no diffusion) is integrated numerically in time using the same average initial conditions as the STZ PDE. 

  The simple ODE model cannot localize and has been used to describe macroscopic frictional behavior for boundary lubrication in thin films~\cite{Anael1} and on earthquake faults~\cite{DaubFV}. The ODE model is an example of a ``rate and state''  friction law.  These laws are frequently used in geophysical modeling of earthquake ruptures, and describe the response of a sheared frictional interface as a function of the slip rate (or strain rate) and a single state variable.  While the STZ PDE resolves internal dynamics of the effective temperature within the interface,  the STZ ODE is constant across the interface.  Comparing the two models allows us to study the effect of small scale dynamics such as strain localization on model predictions for macroscopic behavior.

  In some simulations of the full PDE model, the steady state effective temperature $\hat\chi$ approaches infinity. Although the STZ model given by Eqs.~(\ref{dot-s}) and~(\ref{dot-chi}) is still well-defined in this limit, the shear heating term becomes considerably amplified, indicating a situation where the amorphous packing becomes more and more disordered inside the band.  In every instance where $\hat\chi \to \infty$, the shear band becomes so thin that the numerical integration routine fails.

  We suggest that this numerical failure corresponds to material failure. The smallest length scale in the model is $a$, the diffusion length scale which is on the order of the radius of an STZ. We do not expect the STZ model to hold at length scales smaller than $a$, and because our numerical mesh is fine enough to resolve a band ten times smaller than $a$, numerical failure corresponds to a shear band that rapidly becomes so thin that the model itself breaks down. 

  Although the simple STZ model developed here does not specify the rheology at strain rates above this ``melting'' point, it does suggest that the solid-like STZ theory must be replaced by a liquid-like theory (such as mode coupling or Bagnold scaling) inside these bands.  Therefore, integration of the STZ model indicates that when the disorder temperature approaches infinity, the material can no longer support a static shear stress; it liquefies and fails.

\begin{figure}[h!]
\centering \includegraphics[height=6cm]{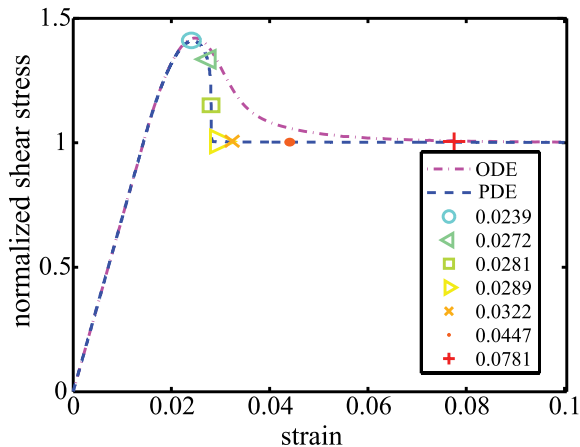}
\caption{\label{fig:RDssdiff} (color online) Shear stress $s$ vs. strain calculated by numerically integrating the STZ equations of motion with initial conditions $\chi_{ini} = 0.0674$, and imposed strain rate $\overline{q} = 1.015 \times 10^{-6}$. The dashed (blue) curve represents the solution to the perturbed STZ PDE model, while the dash-dotted (magenta) curve represents a solution to the STZ ODE model with the same average initial conditions. The colored symbols correspond to the plots shown in Figs.~\ref{fig:RDetdiff} and~\ref{fig:RDsrdiff}. At about 2\% strain, the perturbed system begins to localize and weakens much more rapidly than the magenta curve.}
\end{figure}

  In simulations where the effective temperature remains finite, we numerically track the shear stress $s$ and the effective temperature field $\chi(y)$ as functions of time, or equivalently, strain.  In each case, the stress first responds elastically, and then begins to deform plastically above the yield stress $s_y$.  As plastic deformation increases the effective temperature, the material softens and the stress relaxes to its flowing value, $s_f$.  The dashed blue line in Fig.~\ref{fig:RDssdiff} is a plot of the stress vs. the strain for a numerical solution to the STZ PDE model. Initial conditions are such that the material is highly unstable with respect to shear bands.  For comparison, the dash-dotted (magenta) line in Fig.~\ref{fig:RDssdiff} shows the solution to the STZ ODE model, which is a rate and state law with a single internal state variable. The STZ PDE solution develops a shear band and weakens much more rapidly than a rate and state model with similar initial conditions.
\begin{figure}[h!]
\centering \includegraphics[height=5.5cm]{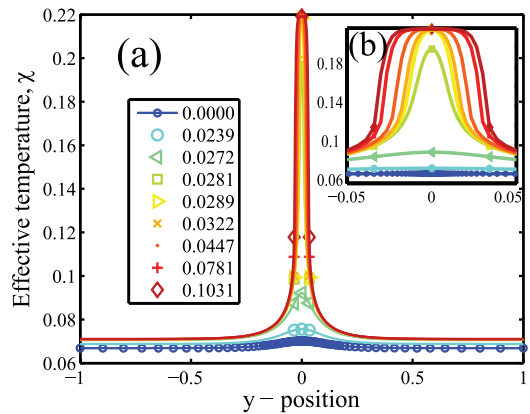}
\caption{\label{fig:RDetdiff}{\bf Diffusion-limited shear band}(color online)  (a) Time series of the effective temperature as a function of position for a material with initial conditions $\chi_{ini} = 0.0674$, and imposed strain rate $\overline{q} = 1.015 10^{-6}$. Each line represents the effective temperature field as a function of position at a different time, as indicated by the legend (all times are in units of strain). The effective temperature field is initially a constant with a small perturbation centered in the middle(blue). This perturbation grows rapidly (green) and forms a shear band, which then diffuses outward slowly (red). (b) Inset shows the same data on a different scale. The strain associated with each line is also indicated in Fig.~\ref{fig:RDssdiff}; localization coincides with rapid dynamic weakening of the shear stress.}
\end{figure}
\begin{figure}[h!]
\centering \includegraphics[height=7cm]{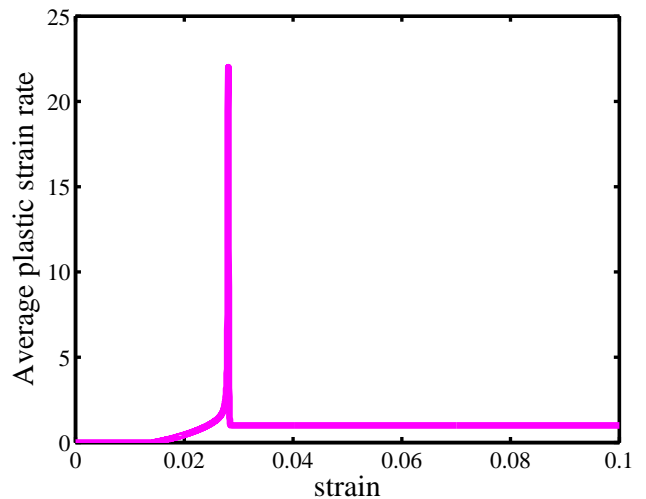}
\caption{\label{fig:RDDpldiff} Average plastic strain rate, $\int dy \; \dot{\gamma} (y) \, \tau_0 / \overline{q}$ as a function of strain for the same integration data shown in Fig.~\ref{fig:RDssdiff}. Initially the average plastic strain rate is zero during the material elastic response.  The average plastic strain rate then rises rapidly during the stress overshoot, when the system releases stored elastic energy. Finally, it relaxes back to unity in the flowing regime, when all the strain must be accommodated plastically.}
\end{figure}
\begin{figure}[h!]
\centering \includegraphics[height=5.5cm]{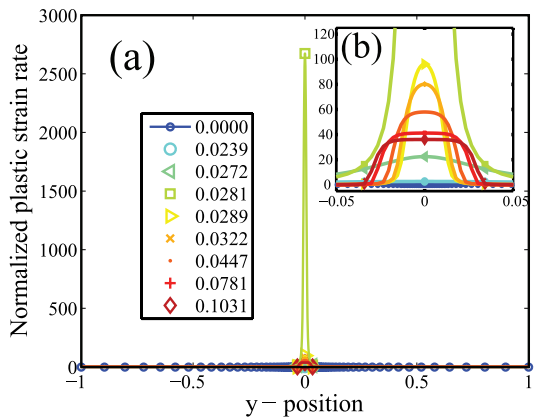}
\caption{\label{fig:RDsrdiff}{\bf Diffusion-limited shear band}(color online)  (a) Time series of the normalized plastic strain rate $\dot\gamma \tau_0 / \overline{q}$ as a function of position.  The plastic strain rate is initially zero (during elastic deformation) but rises at the onset of plastic deformation and becomes very sharply peaked (green). This shear band is extremely narrow with a thickness of about $0.015$, which is approximately the same as the diffusion length scale $a$. As the stress relaxes the strain rate also relaxes, and the shear band becomes wider and less sharply peaked (red).(b)Inset shows magnified position and strain rate axes. Although the maximum strain rate in the band decays significantly with time, it remains large ($> 25$ times the imposed strain rate).}
\end{figure}

  During this initial stress response, the effective temperature field also evolves in space and time.  The effective temperature field is initially constant with a small, centered perturbation, and the field remains static during the elastic response. At the onset of plastic deformation, the effective temperature begins to rise. In systems that deform homogeneously the average value of the effective temperature rises and the perturbation dissipates, while in systems which develop shear bands the effective temperature rises rapidly inside the band and attains a slowly evolving state where the shear bands diffuse outwards.

  Figures~\ref{fig:RDetdiff}(a) and (b) are a series of plots of the effective temperature as a function of position for a material that develops a thin shear band (later we will identify the initial transient as diffusion limited).  Each colored line represents a different time in units of strain. A small initial perturbation to the effective temperature is driven by a dynamic instability to a much higher value, saturating at $\chi \sim 0.22$, and the band then slowly diffuses outward.  
 
  The plastic strain rate also evolves during the initial transient response. We first focus on the {\em average} plastic strain rate, shown in Fig.~\ref{fig:RDDpldiff} as a function of time. At early times when the stress is below the yield stress $s_y$, the system deforms elastically and the plastic strain rate is zero everywhere.  At the onset of plastic deformation the average plastic strain rate increases continuously from zero, attains a maximum, and then relaxes back to the externally imposed strain rate (in the flowing regime all the deformation is plastic.) While the plastic strain rate is greater than unity, stored elastic energy is being dissipated.

  Although the stress is constant across the width of the material, regions with a higher effective temperature deform more rapidly.
To effectively compare strain localization at various strain rates, we plot the dimensionless strain rate, which is the strain rate at each location divided by the externally imposed strain rate. In figures~\ref{fig:RDsrdiff}(a) and (b), each colored line represents a different time in units of strain; plots show the plastic strain rate as a function of position. Initially the stress is below the yield stress and the plastic strain rate is zero (blue line). Localization of the effective temperature field results in a very narrow peak in the strain rate field (green line). The strain rate in the center of the shear band is nearly 3000 times larger than the externally imposed strain rate. As the stress continues to relax, the strain rate becomes less sharply peaked (red line).  The inset plots magnify the position axis.  

Comparing numerical results to analytic predictions requires a method for measuring the degree of localization in a given numerical simulation. The degree of localization can be quantified using the Gini coefficient $\phi$~\cite{Gini}, defined as:
\begin{equation}
\label{giniRD}
\phi(t) = \frac{1}{2 n^2 \overline{D^{pl}}} \sum_i \sum_j \vert \mathcal{ D}^{pl}(y_i,t) -  \mathcal{D}^{pl}(y_j,t) \vert,
\end{equation}
where $\{y_i\}$ is a uniform grid of $n$ points in position space.  The Gini coefficient is equal to zero if the material deforms homogeneously and increases as the plastic strain rate field becomes more sharply peaked. A delta function has a Gini coefficient of 1. During a given numerical simulation, the Gini coefficient $\phi(t)$ starts out as a very small number and then increases rapidly as the shear band forms.  Then, as the shear band diffuses the Gini coefficient decreases. Because we are focusing on the initial transient, we first study the maximum value of the Gini coefficient attained during a given numerical simulation.

\begin{figure}[h!]
\centering \includegraphics[height=12cm]{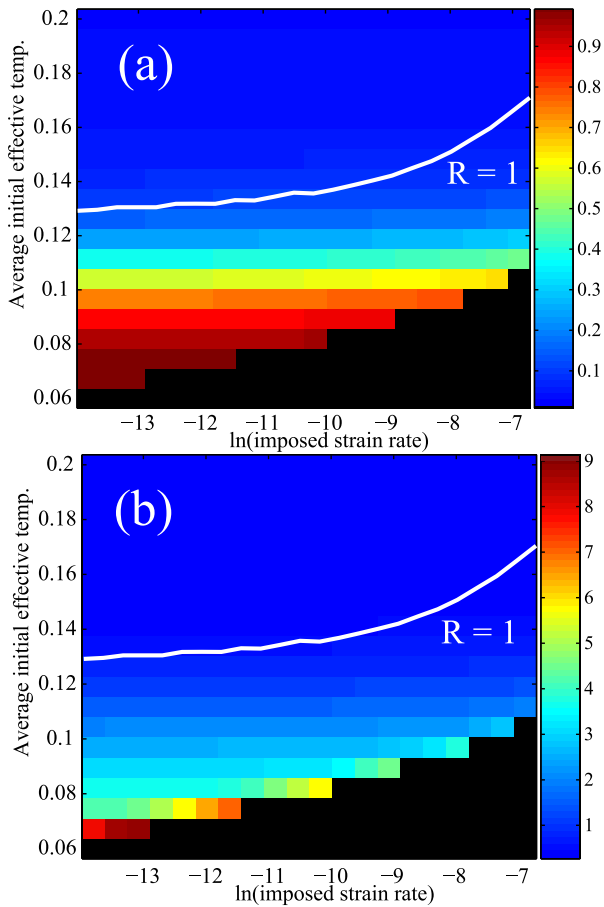}
\caption{\label{fig:RDgini}(color online)  A diagram showing the degree of localization found in by numerically integrating the STZ equations. (a) The maximum Gini coefficient(colorbar), Eq.~(\ref{giniRD}), as a function of the average initial effective temperature and the externally imposed strain rate. A higher Gini coefficient indicates more localization. (b) The log of the maximum plastic strain rate attained in the band(colorbar) divided by the externally imposed strain rate. In both figures, black boxes correspond to numerical simulations where the magnitude of the strain rate was so large that $\hat\chi \to \infty$, as discussed in the text. The solid white line corresponds to the predicted localization ratio $\mathcal{R}= 1$: Localization is expected for initial conditions below this line. See, e.g. results for $\chi_{ini} = 0.0674$, $\overline{q} = 1.015 \times 10^{-6}$ in Fig.4.}
\end{figure}

 Figure~\ref{fig:RDgini}(a) is an intensity plot of the maximum value of the Gini coefficient as a function of the average initial effective temperature, $\chi_{ini}$ and the natural logarithm of the dimensionless imposed strain rate $\log(q)$.  This deformation map indicates that material deformation gradually changes from homogeneous flow to shear banding as a function of the initial conditions.  In figure~\ref{fig:RDgini}(a), black boxes indicate that $\hat\chi$ approached infinity during a particular numerical integration, and the STZ solid-like description breaks down.

  While the Gini coefficient is a direct indicator of localization, it is perhaps a less familiar metric. For comparison, Figure~\ref{fig:RDgini}(b) shows that maximum plastic strain rate attained in the band as a function of the initial conditions. A larger plastic strain rate is attained in a thinner, more localized band, and therefore  Fig.~\ref{fig:RDgini}(b) is very similar to Fig~\ref{fig:RDgini}(a).  Again, black boxes correspond to shear bands where the plastic strain rate reaches the melting point and the model breaks down.

\section{Deformation map and macroscopic implications}
\label{sec:RDlocal_dyn}
 Numerical solutions presented in the previous section show that transient dynamics can lead to inhomogeneous flows. We would like to understand how to characterize these flows.  What type of deformation occurs as a function of the initial conditions? If shear bands form, what sets their thickness? What are the implications of inhomogeneous flows for macroscopic system response?

\subsection{Shear band thickness}

 As mentioned earlier,  the stress appears to achieve a steady state quickly -- after less than 7 \% strain all numerical STZ solutions have acheived a steady stress that changes by less than 5 \% over the course of the remaining simulation (200 \% strain). In comparison, the effective temperature field often remains highly localized for $t > 20 \%$ strain, and broadens over much longer timescales than the stress.

  The goal of this section is to calculate shear band thickness as a function of initial conditions within the STZ model, and determine what sets the thickness of the shear bands in this model. Because localized strain states are transient -- the bands diffuse outward over time -- we study the model predictions for how shear band thickness evolves over large (20 \%) strains. Importantly, many initial conditions lead to numerical simulations and experimental materials that fail before reaching large strains. The following analysis identifies these events as well.

  We calculate the shear band thickness for each numerical solution at two times: the time $t_{qmax}$ at which the strain rate in the shear band attains its peak and the shear band thickness is minimized, and at a later time $t=20 \%$ strain where the stress appears to be in steady state. For systematic study, we specify initial conditions that generate only a single shear band  -- multiple shear bands are often found experimentally at higher strain rates and will be a topic of future study.  

   We first study the shear bands at a time when the plastic strain rate is most highly localized. Let $q_{max}(y)$ be the normalized plastic strain rate $\dot\gamma(y,t) \tau_0 \, / \overline{q}$ evaluated at the time $t_{qmax}$ when the strain rate achieves its absolute maximum. $s_{qmax}$ is the shear stress at $t_{qmax}$.  The thickness $w_{qmax}$ of the shear band in a numerical solution is defined to be the fraction of the real line between $-1$ and $1$ where the function $q_{max}(y)$ is sufficiently large:
\begin{equation}
\label{wqmax}
w_{qmax} = \int_I dy.
\end{equation}
The region $I$ is defined as follows:
\begin{equation}
\label{sqmax}
I = \{y \in [-1,\,1] \; | \;  q_{max}(y) > 1 + h  \sup_y q_{max}(y)\} ,
\end{equation}
where $h$ is an arbitrary fraction. Although we choose $ h = 1/10$, in most cases the calculated thickness is insensitive to the value of $h$ because the strain rate function is sharply peaked. When the system deforms homogeneously, the thickness $w_{qmax}$ is not well-defined. In this case  Eq.~(\ref{wqmax}) becomes extremely sensitive to the fraction $h$ and is no longer accurate.

\begin{figure}[h!]
\centering \includegraphics[height=6cm]{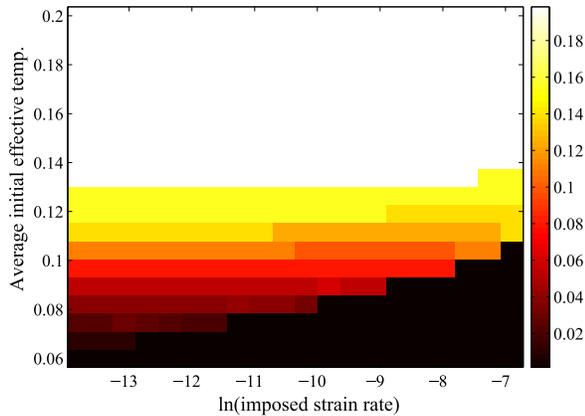}
\caption{\label{fig:RDthickness}(color online)  Shading indicates the shear band thickness at the time of maximum strain rate $w_{qmax}$, Eq.~(\ref{thickness2}), for numerical STZ solutions as a function of the initial conditions $\chi_{ini}$ and log($\overline{q}$). The black boxes correspond to initial conditions for which $\chi \rightarrow \infty$ during an integration, while the white boxes correspond to initial conditions for which the flow is homogeneous (the maximum Gini coefficient, $\phi_{max} < 0.35$) The color scale is set such that the  maximum thickness is 0.2.}
\end{figure}

 Figure~\ref{fig:RDthickness} is a plot of the shear band thickness at the time of maximum strain rate  $w_{qmax}$ as a function of the initial conditions.  White boxes in Figure~\ref{fig:RDthickness} correspond to solutions where the maximum Gini coefficient is less than 0.35. These homogeneously deforming solutions do not have a well-defined value $w_{qmax}$.

\begin{figure}[h!]
\centering \includegraphics[height=6cm]{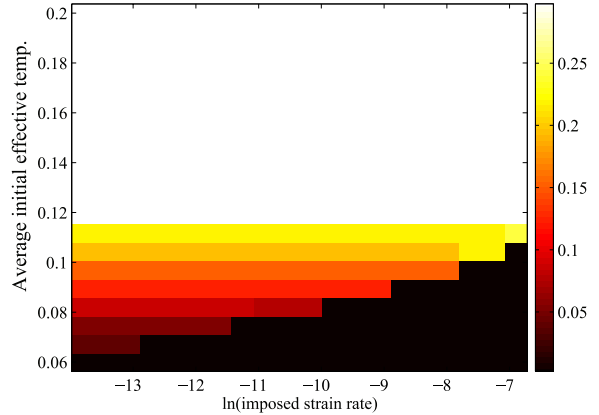}
\caption{\label{fig:RDthickness2}(color online) Shading indicates shear band thickness at 20 \% strain for numerical STZ solutions as a function of the initial conditions $\chi_{ini}$ and log($\overline{q}$). The black boxes correspond to initial conditions for which $\chi \rightarrow \infty$ during an integration, while the white boxes correspond to initial conditions for which the flow is homogeneous (the Gini coefficient at 20 \% strain, $\phi(t= 0.2) < 0.35$) Note that the scale for this plot is larger than that in Figure~7 --  the shear bands are significantly wider at 20 \% strain than at $t_{qmax}$. }
\end{figure}

The discussion in the previous paragraphs analyzes shear bands at their peak, when the plastic strain rate is maximized. This generally occurs at less than 7 \% strain.   Fig.~\ref{fig:RDthickness2} shows the shear band thickness at at 20 \% strain. In each case the shear bands have become wider, as expected. (Note that the maximum thickness shown in this plot is 0.3, as compared to 0.2 in Fig.~\ref{fig:RDthickness}.) Although these systems do not acheive a stationary state, the slowly evolving shear band thickness is observable, and has been seen in molecular dynamics simulations~\cite{Shi} where periodic boundary conditions allow the system to be studied at very large strains.

Perhaps the most interesting feature of Figs.~\ref{fig:RDthickness} and~\ref{fig:RDthickness2} is that the system exhibits no obviously preferred length scale -- the shear band thickness varies continuously from about $\mathcal{O}(a) \simeq 0.015$ to $\mathcal{O}(1)$ (homogeneous flow). Moreover, the shear band thickness increases with time. Both of these observations are a consequence of the fact that localized states are transient solutions to the equations of motion rather than steady state solutions.

  Although there is no preferred shear band length scale, the shear band thickness is reproducible; the STZ model generates shear bands of the same thickness given the same average initial conditions, even if the perturbations are random~\cite{MANNING-LANGER}. In addition, simulations of Lennard Jones glasses generate reproducible shear band thicknesses as a function of time~\cite{Shi} and experiments on bulk metallic glasses find a characteristic shear band thickness~\cite{Lu}. These results suggest that on a given observational time scale, the system does pick out a specific shear band thickness. Because it is observable and reproducible, the shear band thickness must evolve very slowly compared to the stress relaxation time scale. We exploit this feature, showing that the STZ model singles out three different deformation profiles that evolve slowly in time and which should therefore describe observable deformation modes.  In addition, we discuss another state - material failure -- where a parameter in the STZ model diverges and the model fails. 

  We analyze Eq.~\ref{chi-dotRD} to determine what deformation profiles generate the smallest change in the effective temperature. Because the stress is  nearly stationary, flows with the smallest average values for $\dot{\chi}$ are the longest-lived transients and are easily observable. We show deformation profiles for these states and develop a deformation map at  $t_{qmax}$ and $t = 20 \%$ strain. This provides an explanation for observed shear band thicknesses.

\subsection{Relaxation towards homogeneous deformation}

\begin{figure}[h!]
\centering \includegraphics[height=6cm]{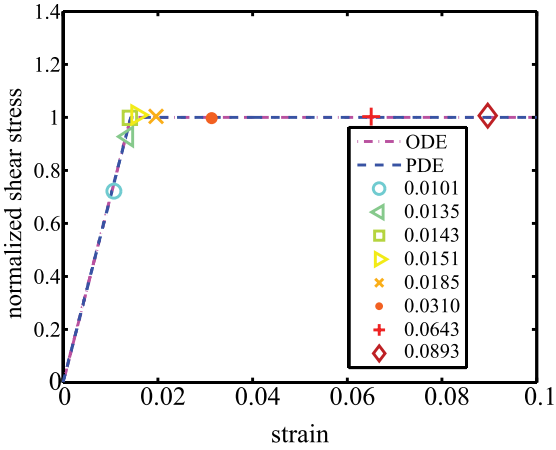}
\caption{\label{fig:RDsshomo}(color online) Deviatoric stress $s$ vs. strain calculated by numerically integrating the STZ equations of motion with initial conditions $\chi_{ini} = 0.0674$, $\overline{q} = 1.015 \times 10^{-6}$. The dashed (blue) curve represents the solution to the perturbed system, while the dash-dotted (magenta) curve represents a homogeneous solution where the effective temperature is constrained to be constant inside the material. In this plot the two curves are indistinguishable. The colored symbols correspond to the plots shown in Fig.~\ref{fig:RDtsrhomo}. Because $\chi_{ini}$ is large, the system begins with a large number of plasticity carriers and therefore the stress peak is negligible. This system does not localize.}
\end{figure}

\begin{figure}[h!]
\centering \includegraphics[height=13cm]{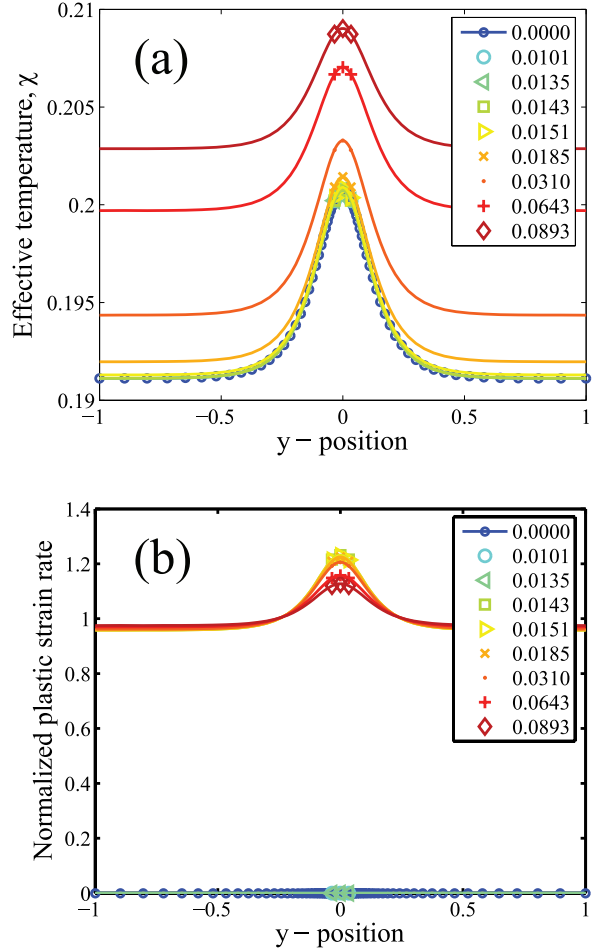}
\caption{\label{fig:RDtsrhomo}(color online)  Relaxation to homogeneous flow: (a) effective temperature and (b) normalized plastic strain rate $\dot\gamma(y) \tau_0 / \overline{q}$ as a function of position, for a material with initial conditions $\int dy\, \chi(y, t=0)= \chi_{ini} = 0.20$, and imposed strain rate $\overline{q} = 1.015 \times 10^{-6}$. Different (colored) lines represent different times. A small initial perturbation to the effective temperature dissipates, although the average effective temperature increases. (The effective temperature scale is much smaller than Fig.~\ref{fig:RDetdiff}(a)). Initially the deformation is purely elastic and the plastic deformation is zero, and at the onset of plastic deformation the average plastic strain rate increases rapidly. Although the plastic strain rate is perturbed at this point, the perturbation decays.}
\end{figure}

 The first and simplest state minimizes $\dot{\chi}$ everywhere and is ``homogeneous deformation.'' The effective temperature field is constant everywhere and equal to $\hat{\chi}(\overline{q})$, where $\overline{q}$ is the externally imposed dimensionless strain rate.  Since both the shear heating and diffusion terms are zero in Eq.~(\ref{chi-dotRD}), this is a true steady state that persists forever.  The dashed blue line in Fig.~\ref{fig:RDsshomo} is a plot of the stress as a function of time for the full STZ model with a small initial perturbation to the effective temperature field, but the initial conditions are such that the deformation relaxes towards homogeneous flow.   The simple ODE model stress solution, shown in magenta, lies on top of the PDE stress solution -- the macroscopic stress response is the same for both models. The colored symbols correspond to plots in Figs.~\ref{fig:RDtsrhomo}(a) and~\ref{fig:RDtsrhomo}(b).

  Homogeneous deformation is characterized by the dissipation of perturbations to the effective temperature field. The effective temperature as a function of position is shown in Fig.~\ref{fig:RDtsrhomo}(a), and each colored line represents the state of the system a different time. A small initial perturbation to the effective temperature dissipates as a function of time, although the {\em average} value of the effective temperature increases as plastic work is dissipated. The effective temperature never varies more than 5\% from its average value.

  A similar plot for the plastic strain rate is shown in Fig.~\ref{fig:RDtsrhomo}(b). The plastic strain rate  is zero during the elastic response, and although the perturbation to the effective temperature generates a small perturbation to the strain rate at the onset of plastic deformation, the strain rate relaxes towards a homogeneous state. The maximum plastic strain rate is remains within 20\% of its average value.


\subsection{Diffusion limited shear bands and failure}

 A second slowly evolving state, called ``diffusion limited localization,''  occurs when the shear heating and diffusion terms in Eq.~(\ref{chi-dotRD}) balance.  In this case the effective temperature field is far from its steady state value $\hat{\chi}$ at all points in space, so that the factor $(1 - \chi / \hat{\chi})$ is close to unity and the shear heating term $s \chi / (s_0 c_0)$, which is of order one, balances the diffusion term $a^2$.  The balance is not perfect, and $\dot{\chi}$ is not exactly zero, so the band continues to diffuse slowly outward. 

 This type of deformation is important because it sets the minimum length scale for shear bands $\mathcal{O}(a) \simeq 0.015$, in numerical STZ solutions.  Although a subset of initial conditions generates shear bands that become thinner than this length scale, the local strain rate in these bands becomes so large that the steady state effective temperature $\hat{\chi}$ approaches infinity.  In other words, $s \chi / (s_0 c_0)$ is too large inside the band, the diffusive flux can not balance it, resulting in a runaway heating process.  This failure is not an artifact of our numerical methods; it signifies a break down in the STZ model that occurs when the effective temperature increases without bound.  We associate this runaway process with a third state, the onset of material failure, because the solid-like STZ description fails as the material liquefies.

  Although we can not track the thickness of the band below the grid resolution during this runaway process, we do observe that just prior to failure the effective temperature in these simulations is elevated significantly above its average in a region of thickness $a$. In other words, the diffusion length scale appears to be an upper bound on the size of the region where structural changes occur during these shear failure events.

 Excluding material failure, the thinnest shear bands possible in this model are diffusion limited.  These types of shear band persist for long times in STZ simulations for earthquake faults, where the parameters are highly rate weakening and chosen to reproduce the rate dependence observed in granular fault gouge experiments~\cite{Daub-Manning}.  Diffusion also limits the initial thickness of shear bands in several of the numerical simulations performed in this paper, although these shear bands continue to diffuse outward at larger strains.

 Diffusion limited shear bands can be identified by their narrow thickness, which is of order $a$, although the exact value varies with the stress overshoot and specific heat $c_0$. In this 2D model, we use the term ``thickness'' to refer to the extent of the shear band in the direction orthogonal to the slip plane, which is similar to the meaning of this term in three-dimensional systems~\cite{Johnson3}. Although diffusion of potential energy (and presumably effective temperature) has been seen in simulations~\cite{Shi}, the length scale $a$ associated with this diffusion constant is relatively unconstrained by simulations or experiments. A reasonable postulate is that $a$ is on the same order as the radius of an STZ, or equivalently, a few particle radii. This suggests that diffusion limited shear bands are very narrow.

The stress vs. strain curve for a material that develops a diffusion limited shear band is given by the dashed blue line in Fig.~\ref{fig:RDssdiff}.  The stress weakens very rapidly as the diffusion limited shear band forms.  For comparison, the dash-dotted magenta curve in Fig.~\ref{fig:RDssdiff} is the stress response of the ODE rate and state model with the same average initial conditions. This illustrates that thin shear bands drastically change the macroscopic system response, and that this dynamic weakening is not captured by a single degree of freedom rate and state model.

 As discussed in Section~\ref{sec:RDnum_res}, Figures~\ref{fig:RDetdiff} and ~\ref{fig:RDsrdiff} show the time evolution of a shear band which is initially diffusion-limited.

\subsection{Disorder limited shear band}

\begin{figure}[h!]
\centering \includegraphics[height=6cm]{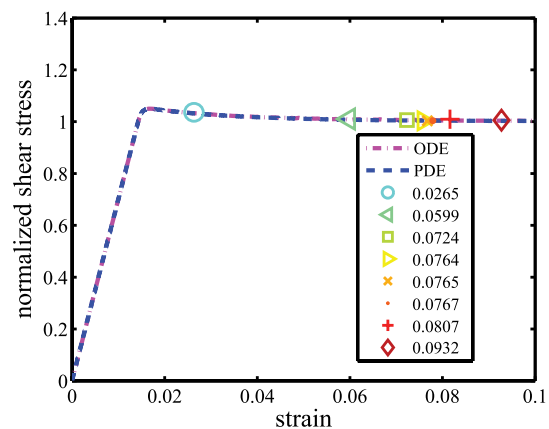}
\caption{\label{fig:RDssdis}(color online)  Shear stress $s$ vs. strain calculated by numerically integrating the STZ equations. The dashed (blue) curve represents the solution to the STZ PDE, while the dash-dotted (magenta) curve represents a solution where the effective temperature is constant inside the material. The colored symbols correspond to the plots shown in Fig.~\ref{fig:RDtsrdis}. Although the localized system weakens slightly faster than the homogeneous system, the effect is small and on this scale the two curves are indistinguishable.}
\end{figure}

\begin{figure}[h!]
\centering \includegraphics[height=12cm]{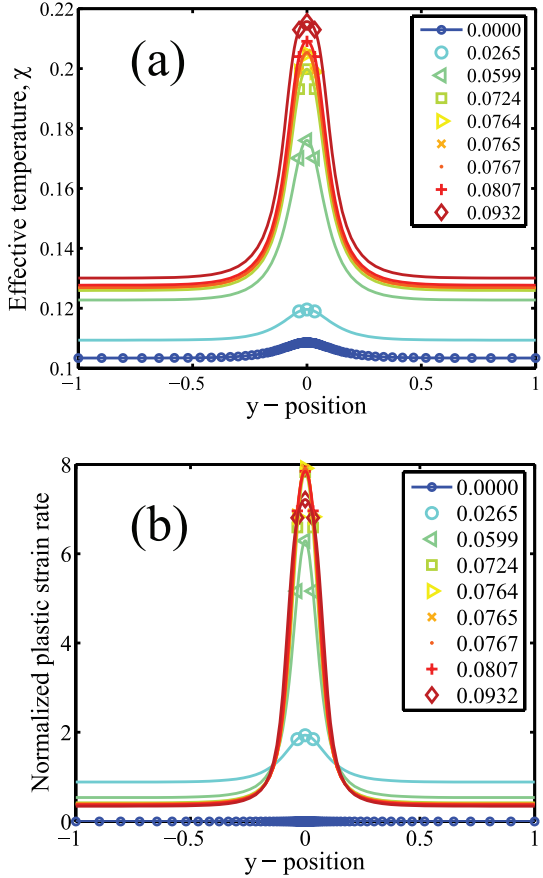}
\caption{\label{fig:RDtsrdis}{\bf Disorder limited shear band}(color online)  (a) Normalized plastic strain rate $\dot\gamma(y) \tau_0 \, / \overline{q}$ and (b) effective temperature as a function of position ($y$), for a material with initial conditions $\chi_{ini} = 0.1042$, and imposed strain rate $\overline{q} = 8.7 \times 10^{-6}$. Different (colored) lines represent different times; cooler colors (blue) correspond to earlier times, while warmer colors (red) correspond to later times. The plastic strain rate in the band increases significantly (about 800 \%), although much less than in the diffusion limited shear band. The thickness of this band at its peak is about 0.2, much larger than the thickness of a diffusion limited shear band. }
\end{figure}

 The fourth ``disorder limited'' localized state is less intuitively obvious, but occurs frequently in our numerical STZ solutions. Neglecting the diffusion term, the right hand side of Eq.~(\ref{chi-dotRD}) is proportional to the product of two factors, $\exp(-1/\chi)$ and $(1 - \chi / \hat{\chi})$. The former is very close to zero whenever $\chi$ is significantly less than $\chi_0$, and the latter is zero when $\chi = \hat{\chi}$.  The disorder limited state occurs exactly when the small-$\chi$ condition is met outside the shear band and $\chi=\hat{\chi}$ inside the band, so that $\dot{\chi}$ in Eq.~(\ref{chi-dotRD}) is always small. However, it is never zero, so that the disorder limited shear bands are also transient solutions that diffuse outward with time.   This type of shear band was first described in~\cite{MANNING-LANGER}, and captures features of shear bands observed in simulations by Shi, {\em et al.}~\cite{Shi}. 
 Figure~\ref{fig:RDssdis} is a plot of the shear stress $s$ vs. strain for a system that develops a disorder-limited shear band. This plot is calculated by numerically integrating the STZ equations of motion with initial conditions $\chi_{ini} = 0.1042$ and $\overline{q} = 8.7 \times 10^{-6}$. The blue curve represents the solution to the perturbed system, while the magenta curve represents a homogeneous solution where the effective temperature is constant as a function of position inside the material. The colored symbols correspond to the plots shown in Fig.~\ref{fig:RDtsrdis}. Although the localized system weakens slightly faster than the homogeneous system, the effect is small and on this scale the two curves are indistinguishable.

 We show the effective temperature and strain rate fields for a numerical solution that exhibits a disorder limited shear band in Figs.~\ref{fig:RDtsrdis}(a) and~\ref{fig:RDtsrdis}(b). The perturbation to the effective temperature field grows very slowly at first, then more rapidly as $\chi \to \hat{\chi}$, and finally the peak begins to diffuse slowly outward. Similarly, the normalized plastic strain rate begins at zero (blue), then rises quickly (green) and relaxes slightly (red).


  The thickness of disorder limited bands is not set by a simple internal length scale such as $a$.  Instead, the thickness is determined dynamically by the externally imposed strain rate and the initial conditions.

  Assume for the moment that a single shear band forms in the material. This is explicitly enforced for the numerical integration in this paper because the initial hyperbolic secant perturbation at $y=0$ leads to a single shear band at that position. In addition, a single shear band is observed in simulations~\cite{Shi} and numerical integration of the STZ model with random perturbations to the initial effective temperature~\cite{MANNING-LANGER} at low strain rates.

 Under this assumption, almost all of the deformation is accommodated in a band of thickness $w$:  
\begin{equation}
\label{oq-w}
\overline{q} = \tau_0 V_0/ L \simeq  \tau_0 \, (w/L) \, \dot{\gamma}_{band}. 
\end{equation}
Using Eqs.~(\ref{RDq-chi}) and~(\ref{oq-w}) we derive the following relationship between the stress $s$, the thickness of the shear band $w$ and the externally imposed strain rate $\overline{q}$:
\begin{equation}
\label{thickness1}
\frac{\overline{q}\,2 L}{w} \simeq 2 f(s) \, \exp \left[- \frac{1}{\hat{\chi}(\overline{q} 2 L / w)}\right].
\end{equation}
  This is not a prediction for the thickness of the shear band, because the final stress, $s$ is not specified.  Unfortunately, we can not derive an approximate value for $s$ because it depends on the entire history of deformation in the material. In addition, the final value of $s$ is generally close to the yield stress, and $f(s)$ is very sensitive to $s$ in this regime.  However, in the next section we will check to see if the shear bands in a given numerical simulation satisfy the criterion given by Eq.~(\ref{thickness1}).

\subsection{Deformation map at the time of maximum deformation rate}

   We now determine which of these states occur and persist as a function of the initial conditions in the numerically integrated solutions.  First, we use the following categories to characterize the deformation at the time $t_{max}$: homogeneous deformation,  diffusion limited shear band,  disorder limited shear band, or material failure. This categorization is somewhat arbitrary because none of the states are stationary; all perturbed states will eventually decay towards homogeneous flow.  However, by identifying these different regions in phase space we hope to identify length scales and features that are observable in experiments. 

 To determine if a shear band thickness is consistent with disorder limited deformation, we rearrange Eq.~(\ref{thickness1}), inserting $w_{qmax}$ and $s_{qmax}$: 
\begin{equation}
\label{thickness2}
\log \left( \frac{\overline{q}\,2  L}{2 f(s_{qmax}) w_{qmax}}  \right) + \frac{1}{\hat{\chi}(\overline{q} 2 L / w_{qmax})} = 0.
\end{equation}

\begin{figure}[h!]
\centering \includegraphics[height=7cm]{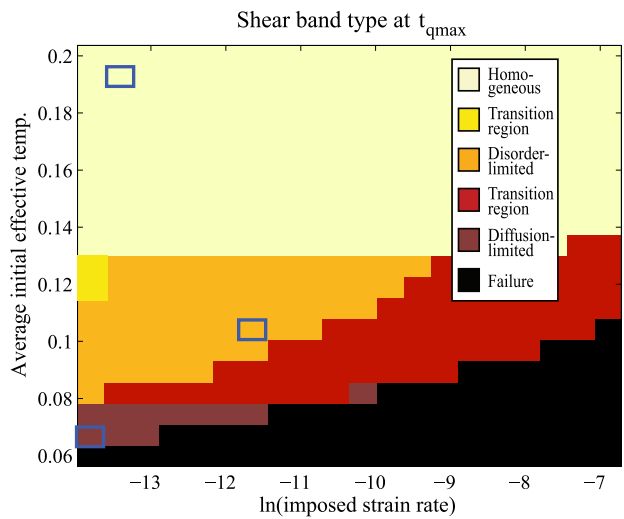}
\caption{\label{fig:RDtype}(color online)  Deformation map that uses the thickness shown in Fig.~\ref{fig:RDthickness} to determine if the deformation at time $t_{qmax}$ is diffusion or disorder limited localization. Diffusion limited shear bands (very dark gray/dark red) and failure (black) occur where $ 0 \leq w < 0.035$, and disorder limited shear bands (medium gray/orange) occur where the left-hand side of Eq.~(\ref{thickness2}) is less than $0.7$.  The very light gray (light yellow) region indicates homogeneous flow. Because this is a snapshot of the the deformation types at $t= t_{qmax}$, for some initial conditions the system is transitioning between two types of flows. The red region represents a transition regime between diffusion limited and disorder limited shear bands, while dark yellow represents a transition between disorder limited shear bands and homogeneous flow. Blue outline boxes indicate initial conditions detailed in Figs.~\ref{fig:RDetdiff},~\ref{fig:RDtsrhomo}, and~\ref{fig:RDtsrdis}.}
\end{figure}

  A shear band in a numerical solution is said to be  ``disorder limited'' if Eq.~(\ref{thickness2}) is satisfied to within 8 \%, (i.e., the left hand side equals $0 \pm 0.08$.)  Similarly, a shear band is ``diffusion limited'' if its thickness is approximately equal to the diffusion length scale $a$, (i.e. $0 < w_{qmax} < 0.03$), and homogeneous if the maximum Gini coefficient is less than $0.5$. These cutoffs are chosen to ensure that deformation regions are non-overlapping, which is a strong constraint. Finally, a material is said to fail if $\hat\chi$ approaches infinity during the course of integration. Transition regions are expected when an inhomogeneous flow does not fit into one of these categories.  Figure~\ref{fig:RDtype} is a deformation map that indicates where each of these criteria are satisfied. 

\subsection{Deformation map at 20 \% strain}
\label{sec:def20}
\begin{figure}[h!]
\centering \includegraphics[height=7cm]{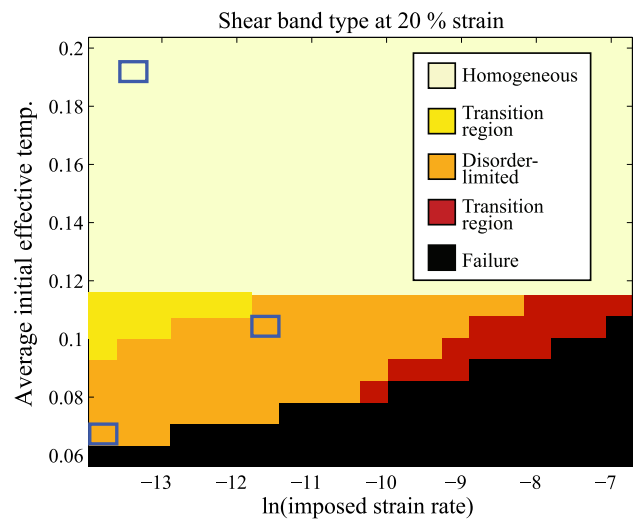}
\caption{\label{fig:RDtype2}(color online)  Deformation map at 20 \% strain, that uses the thicknesss shown in Fig.~\ref{fig:RDthickness2} to determine the type of deformation. Failure (black) occurs where $\hat\chi \to \infty$ during a numerical simulation, and disorder limited shear bands (medium gray/orange) occur where the left-hand side of Eq.~(\ref{thickness2}) is less than $0.7$.  The very light gray (light yellow) region indicates homogeneous flow. At $t = 0.2$, the dark gray (red) transition region in Fig.~\ref{fig:RDtype} has disappeared -- the shear bands have widened to become disorder limited shear bands. In addition, the diffusion limited shear bands have also widened to become disorder limited, and some of the shear bands which were disorder limited in Fig.~\ref{fig:RDtype} have transitioned towards homogeneous flow (light gray/dark yellow). Blue outline boxes indicate initial conditions detailed in Figs.~\ref{fig:RDetdiff},~\ref{fig:RDtsrhomo}, and~\ref{fig:RDtsrdis}.}
\end{figure}

 The same criteria for deformation categories that were used at $t_{qmax}$ in Fig.~\ref{fig:RDtype} can also be used to categorize shear bands at 20\% strain. Inhomogeneous flows which were in a ``transition regime'' at $t_{qmax}$ should broaden quickly towards one of the slowly varying categories. This is seen in Fig.~\ref{fig:RDtype2}, which shows deformation categories at 20 \% strain.  The transition region between disorder limited shear bands and failure has shrunk considerably.  However, all of the diffusion limited flows have transitioned to disorder limited bands, and some disorder limited shear bands have transitioned towards homogeneous flows. This highlights the fact that shear band thickness and deformation type depend significantly on the amount of strain.  Most experiments are limited to small strains and therefore can not see this time evolution.

 This analysis of the STZ model shows that for a large range of initial conditions, shear bands are a robust feature that persist for long times.  STZ theory indicates that shear band thickness evolves slowly over very large strains and suggests that the thickness is determined dynamically by the initial and boundary conditions.  STZ theory predicts that the minimum thickness of the bands is set by an effective temperature diffusion parameter $a$, but a continuum of other thicknesses is also possible and dynamically determined.

 While we predict that the types of deformation mapped in Figs.~\ref{fig:RDtype} and~\ref{fig:RDtype2} will occur in a wide range of amorphous solids, the exact location of boundaries between types and the longevity of each type are likely  material dependent.  These deformation maps depend on the definition of the steady state effective temperature $\hat\chi(q)$. 

 All deformation maps in this work apply specifically to the model glass simulated by Haxton and Liu because we used a function $\hat\chi(q)$ that fits their data for simulated repulsive disks; different materials may have slightly different steady state effective temperatures, although it is also possible that $\hat\chi(q)$ is universal.  The steady state effective temperature can be measured in simulations by comparing the fluctuations and linear response of an observable such as the pressure; this input is all that is needed to generate a deformation map for a new material using STZ theory.

   In the analysis, we have assumed a single shear band.  Experiments and simulations of bulk metallic glasses show that the material develops multiple shear bands at higher strain rates~\cite{Shi, Lu}. Developing a model for the number and spacing between shear bands is beyond the scope of this paper, but the STZ model should provide an excellent starting place for these analyses.

\section{Conclusions}
\label{conclusionsRD}

  We have analyzed the stability of the STZ model with a strain rate dependent effective temperature, and found that the details of the rate dependence specify the steady state stability of homogeneous flows.  Most simulated glasses exhibit rate strengthening, where the steady state stress increases as a function of strain rate, and we have shown that these materials are stable with respect to perturbations in steady state. In contrast, rate weakening materials, such as granular fault gouge, are unstable with respect to shear bands in steady state.

    Perhaps surprisingly, shear bands develop even in rate strengthening materials. They result from an instability that develops during a transient stress response, when a material is driven from rest or driven at a new velocity.  Although the perturbations are unstable only for small strains, the resulting inhomogeneous effective temperature profiles $\chi(y)$ are nearly stationary states of the model equations of motion and therefore these shear bands persist for long times.

 By including information about the rate dependence of the steady state effective temperature, $\hat\chi(q)$, we show that the STZ model generates a deformation map that includes homogeneous deformation, thick ``disorder limited'' shear bands,  thin ``diffusion limited'' shear bands, as well as the onset of material failure. 

  The shear bands that emerge spontaneously in the STZ model capture several important features seen in simulations and experiments. First, the STZ model predicts that shear band formation coincides with stress relaxation after the initial stress overshoot in start-up flows. In cases where the material does not fail, the model predicts that shear bands gradually broaden over large ( $> 20 \%$) strains.

 For a fixed initial effective temperature, the STZ model predicts that the shear bands become thinner and that their internal structure becomes more disordered as the strain rate increases.  This is similar to the ``ductile to brittle'' transition seen in amorphous materials as a function of the strain rate~\cite{Lu}.  At lower strain rates the material deforms nearly homogeneously and appears ductile, but at higher strain rates all the deformation is localized in a thin shear band or mode II crack. 

 At very high strain rates and low initial effective temperatures, the effective temperature approaches infinity at the center of the band during the transient response and the system ``melts''. The liquefied region is very thin and failure occurs near the maximum stress overshoot, which is consistent with material failure via shear banding seen at high strain rates in bulk metallic glasses.

The model predicts that for materials that fail via this shear banding mechanism, the apparent shear band thickness should be at most the diffusion length scale $a$.  Although $a$ has not been measured experimentally, a reasonable assumption is that it is on the order of an STZ radius. In bulk metallic glasses this scale should be at most 30 atomic radii, on the order of 10 nm~\cite{Johnson2}, which is  much  smaller than the thermal diffusion length scale (100-240 nm~\cite{Lewandowski}), and this could explain the shear band thickness measured in these materials.  While the STZ radius has not been estimated in granular fault gouge, this mechanism could provide an explanation for the scale of the prominent fracture surface, which is orders of magnitude smaller than other length scales in earthquake faults.

  We have also shown that these localization dynamics can not be captured by a single degree of freedom rate and state friction law, and that analyzing steady state model dynamics can often be misleading.  This is because the structural degrees of freedom parametrized by $\chi(y)$ evolve much more slowly than the stress dynamics, so that the microstructure continues to evolve although the stress appears to have reached a steady state.  This insight is particularly important for materials that develop highly localized shear bands, as the friction law based on homogeneous dynamics is vastly different from one that accounts for transient shear band development. We suggest that localization may play a role in dynamic weakening seen at high shear speeds in granular materials, and that the STZ PDE model generates a useful friction law in this case. 

  While this is an exciting starting point for studying deformation and failure for amorphous materials, many fundamental questions remain.  We discuss a few of them below.\\

  {\em What is $\hat\chi(q)$ for various amorphous materials?}  Throughout this paper, we have used a fit to data generated by Haxton and Liu~\cite{HL07} as the definition for $\hat\chi(q)$.  Haxton and Liu simulate a 2D amorphous packing of harmonically repulsive discs at thermal temperatures above and below the glass transition temperature, and used FDT to extract an effective temperature at each thermal temperature and strain rate.  To our knowledge, this is the only such data set.  It would be very interesting to use FDT to extract effective temperatures from simulations of other types of amorphous packings, such as the Lennard Jones glass studied by Shi {\em et al.}~\cite{Shi}, foams, amorphous silicon, or bulk metallic glasses. Is $\hat\chi(q)$ similar for all of these materials? Is the effective glass transition temperature, $\chi_0$, universal?

  One possibility is that the transition from glassy behavior to simply activated behavior should occur when $q=1$, (i.e., when the strain rate is the same as the internal rate $1 / \tau_0$).  However, it is also possible that in complicated materials like bulk metallic glasses, the transition occurs at slower rates than $1 / \tau_0$, since the STZs in these systems are large, multi-component regions that likely evolve more slowly than the phonon frequency.

 Are there other ways to measure $\hat\chi(q)$, such as looking at the behavior of a tracer harmonic oscillator inside a simulation box? Is it possible to define the effective temperature by quantifying the change in configurational entropy as a function of the potential energy?  Numerical results from Ono, {\em et al.} suggest that this type of calculation is possible, but they were not able to sample enough low probability states to state conclusively that the FDT and entropic definitions generate the same effective temperature.  These are important questions because it is very difficult to measure fluctuations in position or stress precisely enough in experiments to extract an effective temperature using FDT. \\

  {\em What are the effects of geometry and boundary conditions?} We have so far restricted ourselves to the simplest possible shear geometry and symmetric, no conduction boundary conditions on the effective temperature. The boundary conditions on the effective temperature help determine the location of shear bands within the material as well as the steady states of the system. In many experiments and in some earthquake faults, shear bands tend to localize along the boundary~\cite{Varnik}. Why does this occur?

  Different geometries can be modeled in STZ theory by adjusting the boundary conditions on the effective temperature.  For example, a crystalline solid boundary might impose a constant, more ordered boundary condition on the effective temperature, while a rough, jammed solid might do the opposite. It would be very interesting to investigate the effects of these conditions on shear band evolution.

  In addition, many engineering materials are tested under tension and compression, or a ``notch'' is placed on the surface of the material.  In these cases there is a free boundary which can deform, leading to a coupling between deformation and stress. The necking instability has been investigated using earlier STZ models~\cite{Eastgate} -- it would be interesting to repeat this analysis with our improved understanding of the coupling between structure and deformation.\\

{\em What is the connection between stick slip instabilities and shear banding?}

 In Section~\ref{sec:RDstab}, we showed that the steady-state stability of homogeneous flow was dependent on whether the material was rate strengthening or rate weakening -- rate weakening materials were unstable with respect to shear bands. Interestingly, in rate-and-state (ODE) friction models, stick-slip instabilities can only occur when the system is rate weakening~\cite{Rice}. In addition, formation of a shear band coincides with what looks like a slip event in the macroscopic stress-strain curve. Can a shear band be understood as the PDE analogue of a slip event in a single degree of freedom ODE?  Is is possible for the transient shear bands seen in ostensibly rate strengthening materials to generate stick-slip like behavior?  Preliminary numerical solutions suggest that the transient shear bands can not generate stick slip behavior, but more work is needed on this avenue of research.\\

  {\em  Is the localized state weaker in absolute terms than the homogeneous state? Does this matter?} Recently, researchers studying friction in fault gouge have found that the shear stress supported by the gouge weakens rapidly at high driving rates~\cite{Mizoguchi}.  Strain localization in the STZ model generically leads to a rapid decrease in the shear stress, and has been suggested as a mechanism for this experimental observation~\cite{Daub-Manning}.  However, in most numerical solutions to the STZ equations (e.g. Fig.~\ref{fig:RDssdiff}), the final stress state in the localized system is equal to or higher than the stress in the homogeneous system, except in the special case where the initial effective temperature perturbation is a step function~\cite{Daub-Manning}. In absolute terms, the localized system is stronger (or at least no weaker) than the homogeneous system, which is counter-intuitive.

 There are several ways to reconcile this information with intuition. First, we note that the {\em rate} at which the localized system weakens is much more rapid than the homogeneous system.  For dynamic phenomena, such as stick-slip instabilities and stop-start experiments, the weakening rate and the total stress drop help determine the dynamic response. Is the rapid weakening seen in systems that localize large enough to cause stick-slip?  Another possibility is that many of these systems attain strain rates at which the STZ solid-like description breaks down. Although we do not explicitly model this here, it seems likely that the liquid-like material in the band possess a vastly reduced strength compared to the solid outside the band.  

\begin{acknowledgments}
 This work was supported by the Southern California Earthquake Center, the David and Lucile Packard Foundation, and NSF grant number DMR-0606092. M.L.M. acknowledges an NSF Graduate Research Fellowship. J.S.L. was supported by DOE grant number DE-FG03-99ER45762. 
\end{acknowledgments}

\appendix
\section{STZ model details}
\label{sec:RDmodel_detail}
A mean field theory for shear transformation zones has been developed in a series of papers~\cite{Falk_L1, Falk_L2,Langer_Eff, Bouchbinder}, and we use this theory as a general model for a wide range of amorphous solids. 

   In analyzing the dynamics of shear transformation zones, we develop equations of motion for five internal variables: the deviatoric stress $s$, the pressure $p$, the density of STZs oriented parallel and perpendicular to the principal stress directions $n_{\pm}$, and the effective temperature, $\chi$. In a simple shear geometry at low temperatures, the model can be further simplified so that the state of the system is entirely specified by $s$ and $\chi$ alone. The following sections review the STZ equations and specify the parameters and simplifications used in this paper.

\subsection{Overview of equations of motion}
 In the slowly sheared materials we are modeling, the speed of sound in the material is very fast compared to the rate of plastic deformation. In this case the stress gradients equilibrate very quickly, and we take the zero density limit of the momentum conservation equations. This results in static elastic equations for the stress:

\begin{equation}
\label{stat_eq}
\frac{\partial \sigma_{i j}}{\partial x_{j}} = 0.
\end{equation}
The rate of deformation tensor is the sum of elastic and plastic parts:
\begin{eqnarray}
\label{Dtotal}
D^{total}_{i j} &=& \frac{1}{2} \left( \frac{\partial v_i}{\partial x_j} + 
 \frac{\partial v_j}{\partial x_i} \right)  \nonumber \\
 &=& \frac{\mathcal{D}}{\mathcal{D} t} \left( -\frac{p}{2 K} \delta_{i j} + \frac{s_y}{2 \mu} s_{i j} \right) + D^{plast}_{i j},
\end{eqnarray}
 where $\mathcal{D} / \mathcal{D} t $ is the material or co-rotational derivative. To simplify notation, the deviatoric stress has been nondimensionalized by an effective shear modulus $s_y$ that specifies the stiffness of the STZs. The stress scale $s_y$ also characterizes the stress at which the material begins to deform plastically. This yield stress is distinct from the maximum stress attained, $s_m$, and the steady state flow stress, $s_f$, both of which are sometimes also referred to as the yield stress in the literature.

The plastic rate of deformation tensor can be written in terms of dynamical variables from STZ theory. We postulate that under shear stress, each STZ deforms to accommodate a certain amount of shear strain, and cannot deform further in the same direction. This is modeled by requiring that each STZ be in one of two states: oriented along the principal stress axis in the direction of applied shear, which we will denote ``$+$'', or in the perpendicular direction, ``$-$''. 

 Under applied strain, the STZ will {\em flip} in the direction of strain, from ``$-$'' to ``$+$''. Under shear stress in the opposite direction, the STZs can revert to their original configurations, which corresponds to a flip from ``$+$'' to ``$-$''. We assume that the STZ density is small and each STZ interacts with other STZs through continuum fields such as the stress. Therefore the rearrangements or {\em flips} occur at a rate $R(s)/\tau_0$, which depends on the stress and a characteristic attempt frequency $1/\tau_0$.

  Because each STZ can flip at most once in the direction of applied strain, STZs must be created and annihilated to sustain plastic flow.   Based on these considerations, the number density of STZs in each direction, $n_{\pm}$, obeys the following differential equation
\begin{equation}
\label{number_density}
\tau_0 \dot{n}_{\pm} = R(\pm s) n_{\mp} - R(\mp s) n_{\pm} + \Gamma \left( \frac{n_{\infty}}{2} e^{-1/ \chi} - n_{\pm} \right), 
\end{equation}
where $R(\pm s)/ \tau_0$ is the rate of switching per STZ as a function of stress, $\Gamma $ is the rate at which energy is dissipated per STZ, and $n_{\infty}\,  e^{-1/ \chi}$ is the steady state density of STZs in equilibrium. 

 The plastic rate of deformation tensor is given by the rate at which STZs flip:
\begin{equation}
\label{Dpl_R}
D^{pl} = \frac{\epsilon_0}{n_{\infty}\tau_0} \left( R(s) n_{-} - R(-s) n_{+} \right),
\end{equation}
 where $\epsilon_0$ is a strain increment of order unity and $n_{\infty}$ is a density roughly equal to the inverse of the volume per particle.

The first two terms in Eq.~(\ref{number_density}) correspond to STZs switching from ``$+$'' to ``$-$'' states and vice-versa, while the last term enforces detailed balance: STZs are created at a rate proportional to $n_{\infty} e^{-1/ \chi}$ and annihilated at a rate proportional to their density. The creation rate is proportional to the probability of a configurational fluctuation that corresponds to an STZ. As discussed in the introduction, this probability is $\exp[-1/ \chi]$, where $\chi$ is an internal state variable that characterizes the configurational disorder. To close the system of equations, the model requires an equation of motion for $\chi$.

 Ono {\em et al.}~\cite{Ono} and Haxton and Liu~\cite{HL07} show that a driven amorphous system possesses a well-defined steady state effective temperature, $\hat{\chi}$, at each value of the imposed strain rate. In these simulations, a thermostat ensures homogeneous deformation within the glass and the particles are sheared for long periods of time before the steady state measurement is taken.

 To model deformation in time varying systems, such as start-up flows, we have to estimate how the effective temperature changes in time.  As detailed in~\cite{JSL08}, we assume that the heat content in the configurational degrees of freedom is driven by two independent sources, mechanical work and thermal fluctuations.  Therefore we make the simplest assumption: the mechanical heat drives the effective temperature towards $\hat\chi$ according to the conventional linear law of heating. The rate of heat per unit volume that enters the configurational degrees of freedom is $Q_c = T_{eff} \left(d S_c / dt \right)_{mech}$.

 In addition, we postulate that the heat produced by thermal fluctuations, $Q_T = T_{eff}\left(d S_c / dt \right)_{therm}$, drives the effective temperature towards the thermal equilibrium bath temperature $\theta$ according to the linear law of cooling.  The resulting equation of motion for $\chi$ is:
\begin{eqnarray}
\label{chi-sI}
 \dot{\chi} &=&\frac{1}{C_{ef\!f} T_z} \left[ T_{eff} \left(\frac{dS_c}{dt}\right)_{mech} \left[ 1 - \frac{\chi}{\hat\chi} \right] \right. \nonumber \\
 & &\left. + \, T_{eff} \left(\frac{dS_c}{dt}\right)_{therm} \left[ 1 - \frac{\chi T_z}{T} \right] \right] +  D \frac{\partial^2 \chi}{\partial y^2},
\end{eqnarray}
where $C_{ef\!f}$ is a specific heat, $T_z = E_z/ k_B$ is the STZ formation energy in temperature units, and the last term represents diffusion of effective temperature.

 Because the effective temperature governs the configurational degrees of freedom, only configurational rearrangements, i.e. plastic events, permit diffusion of the effective temperature. This suggests that the diffusivity should vary with strain rate, so that  $D = a^{2} \vert \dot{\gamma}_{pl} \vert$, where $a$ is a length scale that corresponds to the radius of an STZ.

\subsection{Simplifying assumptions}

Pechenik~\cite{Pechenik} generalized Eqs.~(\ref{number_density})~and~(\ref{Dpl_R}) to the case where the principal axes of the STZ orientation tensor $n_{ij}$ are not aligned with principal axes of the stress tensor $s_{ij}$.  These generalized equations can be written in terms of two new variables, $\Lambda$ and $m$, which appear often in literature on STZs:
\begin{eqnarray}
\label{Ldef}
\Lambda &\equiv& n_{tot} / n_{\infty}; \\
\label{mdef}
 m_{ij} &\equiv& n_{ij} / n_{\infty},
\end{eqnarray}
 where $n_{tot}$ is the tensorial generalization of ($n_{+} + n_{-}$) and $n_{ij}$ is the tensorial generalization of ($n_{+} - n_{-}$). The scalar $\Lambda$ is the total density of zones in a sample, while the tensor $m$ corresponds to the STZ orientational bias.

  In this paper we focus on materials in a 2D simple shear geometry, so that the diagonal terms in the deviatoric stress tensor ($s_{xx}, s_{yy}$) and STZ orientational bias ($m_{xx}, m_{yy}$) are significantly smaller than off-diagonal terms and can be neglected. Let $s = s_{xy} = s_{yx}$ and $m = m_{xy} = m_{yx}$. In this geometry the pressure $p$ does not change with time and $\mathcal{D}p/\mathcal{D}t$ in Eq.~(\ref{Dtotal}) is zero.

As noted in~\cite{Bouchbinder} the density of STZs, $\epsilon_0 \Lambda$, is necessarily small.  In a simple shear geometry,  the equations of motion for the stress $s$ and the effective temperature $\chi$ each contain this factor in their numerators, and they equilibrate very slowly compared to $m$ and $\Lambda$.  Therefore we replace $\Lambda$ and $m$ by their steady state values. Combining Eqs.~(\ref{number_density}),~(\ref{Ldef}),~and~(\ref{mdef}), we find that the steady state value of $\Lambda$ is $\exp[-1/\chi]$, and that $m$ exchanges between two steady states (elastic vs. plastic deformation) near $s = s_y$.  Below the yield stress the deformation is almost entirely elastic because all the existing STZs are already flipped in the direction of stress. Above the yield stress STZs are continuously created and annihilated to sustain plastic flow.  Details can be found elsewhere~\cite{JSL08}.

  We also make the simplifying assumption that the material is below the thermal glass transition temperature $T_0$, so that particle rearrangements are not activated by thermal fluctuations. In this case the thermal entropy contribution to the effective temperature equation of motion,  $\left(d S_c / dt \right)_{therm}$ in Eq.~(\ref{chi-sI}), is zero. This is always true, for example, in granular materials.  In addition, we make the approximation that at very low temperatures the STZs do not flip in a direction opposite the direction of applied stress: $R(-|s|) = 0$.

  The functional form of $\Gamma$, the energy dissipated per STZ that appears in Eq.~(\ref{number_density}), is considerably simplified for $T < T_0$. Under the assumption that an STZ does not flip in a direction opposite to the direction of applied stress, no energy can be stored in the plastic degrees of freedom. This means that the plastic work is equal to the energy dissipated:
\begin{equation}
\label{plsWork}
D_{ij}^{pl} s_{ij}= \dot \gamma s = Q.
\end{equation}
 Following Pechenik~\cite{Pechenik}, we postulate that the total energy dissipation rate, $Q$, is proportional to $\Gamma$:
\begin{equation}
\label{Q-G}
Q = s_0 \frac{\epsilon_0}{\tau_0} \, \Lambda \, \Gamma ,
\end{equation}
where $s_0$ is a stress scale we return to below.

Combining Eqs.~\ref{Dpl_R},~\ref{plsWork}, and~\ref{Q-G} results in the following expression for $\Gamma$: 
\begin{equation}
\label{GammanoT}
\Gamma(s) = \frac{2}{s_0 \epsilon_0}\,s\,f(s),
\end{equation}
 where the function  $f(s)$, which also appears in Eq.~(\ref{RDq-chi}), is defined as follows:
\begin{equation}
\label{f-sI}
f(s) = \epsilon_0 \frac{\mathcal{R}(s)}{2} \Bigl[1 - m(s)\Bigr].
\end{equation}

In simple shear below the thermal glass transition, the two steady states of the equation of motion for the STZ bias are also very simple:
\begin{equation}
m(s) \to \begin{cases} 1 & \mbox{for}  \; s < s_y \\ 
   s_0 / \,s & \mbox{for} \; s > s_y, \end{cases}
\end{equation}
and $s_0 = s_y$ is the yield stress -- the stress at which the two stability branches intersect. 
 
   We now turn to the parameters in Eq.~(\ref{chi-sI}). In~\cite{JSL08} the rate at which configurational entropy is being produced by mechanical deformation, $(d S_c / dt)_{mech}$, is assumed to be proportional to the product of the STZ density and the STZ creation rate:
\begin{equation}
\label{entForm}
\left( \frac{d S_c}{dt} \right)_{mech} = \frac{k_B \nu_Z}{\Omega}  \frac{\epsilon_0}{\tau_0} \Lambda \Gamma, 
\end{equation}
where $\Omega$ is a volume per molecule and $k_B \nu_Z$ is the entropy per STZ~\cite{JSL08}.

  Putting everything together, we arrive at an equation of motion for the effective temperature:
\begin{eqnarray}
\label{dot-chiA}
\frac{ d \chi}{d \gamma} &=& \frac{2  \: s \chi}{\tilde{c}_0 s_0 \overline{q}}f(s)  e^{-1/ \chi}  ( 1 - \frac{\chi}{\hat{\chi}(q)}) + \:  a^2 \dot{\gamma}_{pl}  \frac{\partial^2 \chi}{\partial y^2},
\end{eqnarray}
where $\gamma$ is strain, $\overline{q}$ is the imposed strain rate times the STZ time scale, $(V_0/ L) \, \tau_0$ and $\tilde{c}_0 = C_{eff} \Omega / \,(k_B \nu_{z}) $.  Inserting Eq.~(\ref{RDq-chi}) into the equation for the rate of deformation tensor, Eq.~(\ref{Dtotal}), and integrating across the width of the material (in the $y$-direction) results in a second equation for the stress dynamics: 
\begin{equation}
\label{dot-sA}
\frac{ds}{d \gamma} = \mu_{*} \left( 1 - \frac{2 }{\overline{q}} \, f(s)\, \overline{\Lambda} \right),
\end{equation}
where $\mu_{*}$ is the ratio of the elastic modulus to the yield stress, and $\overline{\Lambda}$ is the spatial average of the STZ density $\Lambda = \exp(-1/ \chi)$. 

   Determining an exact function $R(s)$ from first principles is a difficult many body problem. However, we do know how $R(s)$ behaves in the limits of very small and very large stresses, and we choose a function that smoothly interpolates between these two regimes.  $R(s)$ exhibits Eyring-like behavior far below the yield stress, and power law behavior above the yield stress:
\begin{equation}
\label{R-sSTZ}
R(s) = \exp \left[-{T_E\over T}\,e^{-s/\tilde\mu}\right]\left[1 + \left({s\over s_1}\right)^{2} \right]^{n/2}.
\end{equation}
The first factor on the right-hand side of Eq.(\ref{R-sSTZ}) is the Eyring rate in a form similar to that used in \cite{Falk_L1}, where the exponential function of $s/\tilde\mu$ causes the rate to saturate at large $s$.  Here, $T_E$ is the height of the Eyring activation barrier in units of temperature.   The exponent $n$ in Eq.(\ref{R-sSTZ}) specifies the large stress power law behavior; possible values are discussed in~\cite{LANGER-MANNING}.  Analysis of simulation and experimental data suggest that $n=1$ is valid for bulk metallic glasses~\cite{JSL08} but in this paper we use $n=1/2$, which is relevant for purely repulsive harmonic disks~\cite{LANGER-MANNING}.

\end{document}